\documentclass[aps,prb,showpacs,floatfix,reprint]{revtex4-1}

\usepackage{graphicx}
\usepackage{float}
\usepackage{color}
\usepackage{amsmath,amssymb}
\usepackage{bm}
\usepackage{upgreek}

\begin{document}

\title{Probing magnetism via spin dynamics in graphene/2D-ferromagnet heterostructures}
\author{Aron W. Cummings}
\email{aron.cummings@icn2.cat}
\affiliation
{
	Catalan Institute of Nanoscience and Nanotechnology (ICN2), CSIC and BIST,
	Campus UAB, Bellaterra, 08193 Barcelona, Spain
}
\date{\today}

\begin{abstract}
The recent discovery of two-dimensional magnetic insulators has generated a great deal of excitement over their potential for nanoscale manipulation of spin or magnetism. One intriguing use for these materials is to put them in contact with graphene, with the goal of making graphene magnetic while maintaining its unique electronic properties. Such a system could prove useful in applications such as magnetic memories, or could serve as a host for exotic states of matter. Proximity to a magnetic insulator will alter the spin transport properties of graphene, and the strength of this interaction can be probed with Hanle spin precession experiments. To aid in the analysis of such experiments, in this work we derive an explicit expression for Hanle spin precession in graphene interfaced with a ferromagnetic insulator whose magnetization points perpendicular to the graphene plane. We find that this interface results in a shifted and asymmetric Hanle response, and we discuss how this behavior can be used to interpret measurements of spin transport in these systems.
\end{abstract}

\maketitle

\section{Introduction}

Since its isolation and measurement 15 years ago \cite{Novoselov2004}, graphene has emerged as a highly promising material for a wide range of applications, a result of its unique electrical, thermal, optical, mechanical, and chemical properties \cite{Ferrari2015}. Graphene also shows great promise for the field of spintronics, which aims to use an electron's spin, instead of its charge, as a means of carrying and manipulating information \cite{Roche2015}. In particular, owing to its high electron mobility, small spin-orbit coupling (SOC), and negligible hyperfine interaction, graphene has proven to be an extremely efficient transporter of spins, with measured spin relaxation lengths in the range of tens to hundreds of $\upmu$m \cite{Dlubak2012, Drogeler2016}.

However, these same features that make graphene an optimal spin conductor, namely its small SOC and nonexistent magnetism, make it ineffective for the active generation or manipulation of spins and spin currents. To this end, recent work has focused on interfacing graphene with insulating materials that have large SOC, such as transition metal dichalcogenides (TMDCs) or topological insulators (TIs), in the hope that their proximity will induce strong SOC in graphene while maintaining its superior charge transport properties.

With respect to graphene/TMDC heterostructures, this approach has proven to be highly successful: measurements of weak antilocalization (WAL) indicate enhanced SOC in the graphene layer \cite{Wang2015a, Wang2016, Yang2016, Yang2017, Volkl2017, Wakamura2018, Zihlmann2018}; spin switches have been realized based on spin absorption in the TMDC layer \cite{Yan2016, Dankert2017}; predictions of giant spin relaxation anisotropy \cite{Cummings2017}, which may be useful for orientation-dependent spin filtering, have been subsequently confirmed by experiments \cite{Ghiasi2017, Benitez2018}; and recent measurements have indicated sizable charge-to-spin conversion in the graphene layer \cite{Safeer2019, Ghiasi2019}. Giant spin lifetime anisotropy has also been predicted in graphene/TI heterostructures \cite{Song2018}, and recent measurements of spin transport and WAL have suggested that TIs also induce strong SOC in graphene \cite{Khokhriakov2018, Jafarpisheh2018}.

In order to realize exotic states of matter such as the quantum anomalous Hall effect in graphene, it is necessary to combine SOC with a magnetic exchange field that is perpendicular to the graphene plane \cite{Qiao2010}. To this end, initial efforts were made to induce magnetism in graphene by interfacing it with yttrium iron garnet (YIG), an insulator that is ferromagnetic up to 560 K \cite{Anderson1964}. Charge transport measurements have suggested the presence of a perpendicular exchange field induced in the graphene layer via proximity to YIG \cite{Wang2015b, Mendes2015, Tang2018}. However, spin transport measurements indicate that while in-plane ferromagnetism can be induced in the graphene layer, the out-of-plane exchange field remains purely paramagnetic and thus disappears in the absence of an external magnetic field \cite{Leutenantsmeyer2016}. This problem may be overcome with the recent discovery of various two-dimensional ferromagnets (2DFMs) such as CrI$_3$, Cr$_2$Ge$_2$Te$_6$ (CGT), or CrBr$_3$, which can exhibit perpendicular magnetic anisotropy \cite{Huang2017, Gong2017, Ghazaryan2018}. Indeed, recent measurements of spin transport in graphene/CGT heterostructures show hysteresis in the Hanle spin precession, indicative of a perpendicular proximity-induced exchange field at zero external magnetic field \cite{Karpiak2019}. These measurements reveal the possibility of using proximity effects to induce out-of-plane ferromagnetism in graphene, and to possibly combine them with SOC proximity effects to realize more exotic phases of matter.

Given these recent measurements and the current interest in graphene and 2DFMs, it is useful to have an explicit model describing the nature of spin transport in heterostructures of these materials. In this work we derive a model that describes Hanle spin precession in graphene/2DFM heterostructures, in the case where the 2DFM induces an out-of-plane exchange field in the graphene layer. We account for the nonuniformity of the graphene channel, which allows one to separately consider the nature of spin transport in the pristine graphene regions and in the region where graphene is interfaced with the 2DFM. Our model reveals two characteristic behaviors: 1) the local exchange field induced by the 2DFM shifts the peak of the Hanle curve, with the magnitude of this shift proportional to, but not necessarily equal to, the strength of the exchange field; and 2) the nonuniformity of the proximity-induced exchange field causes the Hanle curve to become asymmetric with respect to its maximum. After deriving the model in Sec.\ \ref{sec_model}, in Sec.\ \ref{sec_model_features} we show how the shape of Hanle curve depends on various parameters. Finally, in Sec.\ \ref{sec_spurious} we show how spurious contact effects can give qualitatively the same behavior as a proximity-induced exchange field, and we discuss approaches to distinguish the two. Overall, we hope our derived model and the ensuing discussion will prove useful for the analysis of future measurements of spin transport in graphene/2DFM heterostructures.

\section{The Model} \label{sec_model}

\begin{figure}[t]
\includegraphics[width=\columnwidth]{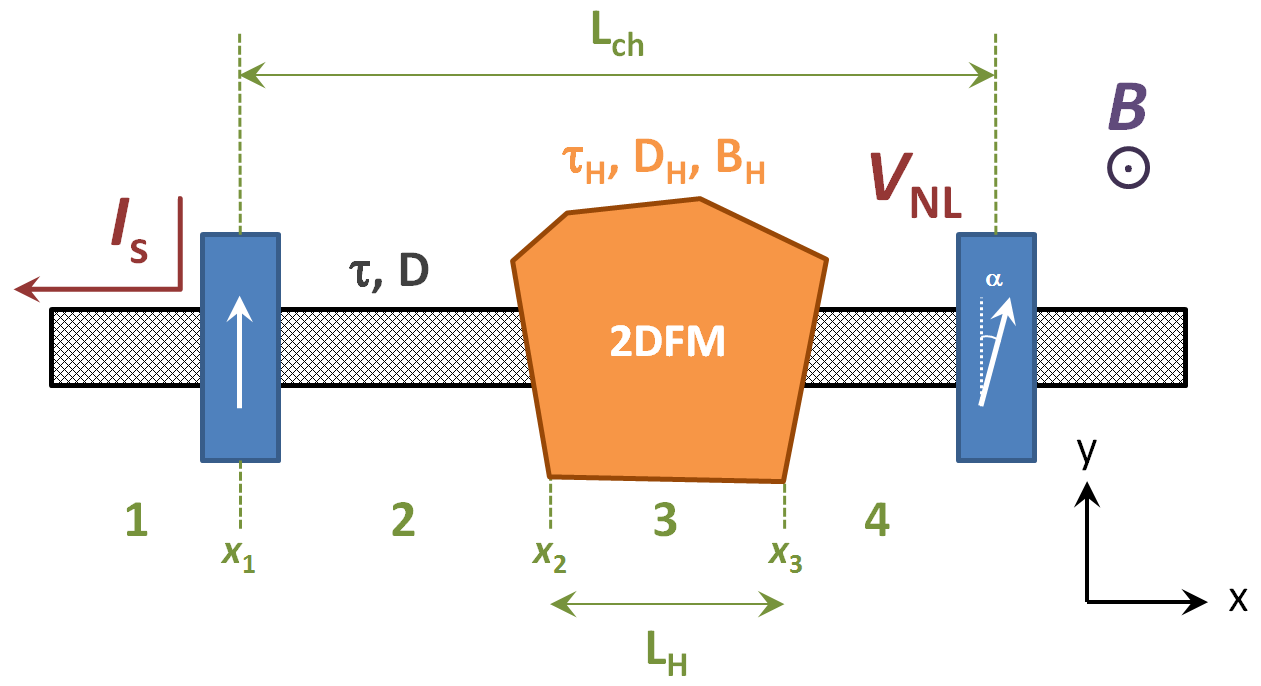}
\caption{Typical nonlocal spin valve measurement setup. A spin current $I_\text{s}$ is injected from a ferromagnetic contact to the left, a buildup of spins $\mathbf{s}$ diffuses through the graphene to the right, and is measured as a nonlocal voltage $V_\text{NL}$. An external perpendicular magnetic field $B$ induces precession of the spin in the graphene plane. A piece of 2DFM placed on the graphene channel induces a local out-of-plane exchange field $B_\text{H}$. Additionally, the spin transport characteristics in the graphene/2DFM region ($\tau_\text{H}, D_\text{H}$) can be different from the pristine graphene region ($\tau, D$). Finally, the injector and detector contacts can be misaligned in the $x$-$y$ plane by an angle $\alpha$.}
\label{fig_nonlocal}
\end{figure}

To describe spin transport in graphene, we consider a typical nonlocal spin valve measurement, as depicted in Fig.\ \ref{fig_nonlocal}. In this setup, ferromagnetic contacts are deposited on top of graphene. The magnetizations of these contacts are typically aligned in the graphene plane and perpendicular to the direction of transport, i.e., along the $y$-axis in Fig.\ \ref{fig_nonlocal}. A spin current $I_\text{s} = P_\text{i}I/e$ is driven from the injector contact to the left side of the device, where $I$ is the charge current driven through the injector contact, $P_\text{i}$ is its spin injection efficiency, and $e$ is the electron charge. This results in a density $\mathbf{s}$ of spin-polarized electrons below the injector. This buildup of spin density will then diffuse away from the injector contact, and those that reach the detector contact can be measured as a nonlocal voltage $V_\text{NL}$. The magnitude of this voltage is proportional to the density of spins that reach the detector polarized along its magnetization axis.

To study the dynamics and relaxation of the spins as they diffuse through the graphene, a magnetic field $\mathbf{B}$ is applied perpendicular to the graphene plane. The dependence of the nonlocal voltage $V_\text{NL}$ on the strength of the magnetic field, known as a Hanle curve, can be used to extract such parameters as the spin lifetime and the spin diffusion length in the graphene channel. In this section we will derive an analytical expression for the Hanle curve, which can then be fit to experiments.

\subsection{General solution of the spin diffusion equation}

The starting point of our analysis is the spin diffusion equation,
\begin{align} \label{eq_spin_diff_gen}
\frac{\partial \mathbf{s}}{\partial t} = D \frac{\partial^2 \mathbf{s}}{\partial x^2} - \Gamma\mathbf{s} + \omega (\mathbf{s} \times \hat{e}_B),
\end{align}
where $\mathbf{s}(x,t)$ is the spatially- and time-dependent spin density, whose vector nature describes the orientation of the spin polarization. The motion of the spins is captured by the spin diffusion coefficient $D$. The last term describes spin precession arising from the applied magnetic field $\mathbf{B}$, where $\omega = \gamma |\mathbf{B}|$ is the Larmor precession frequency, $\gamma$ is the gyromagnetic ratio, and $\hat{e}_B$ is the unit vector denoting the orientation of $\mathbf{B}$. Spin relaxation is captured by the matrix $\Gamma$,
\begin{align}
\Gamma = 
\begin{bmatrix}
1/\tau_\text{s}^x & 0 & 0 \\
0 & 1/\tau_\text{s}^y & 0 \\
0 & 0 & 1/\tau_\text{s}^z
\end{bmatrix}, \nonumber
\end{align}
where $\tau_\text{s}^i$ is the lifetime of spins oriented along axis $i$.

To solve Eq.\ (\ref{eq_spin_diff_gen}), we assume a magnetic field perpendicular to the graphene plane, $\mathbf{B} = B\hat{z}$, and we look for a steady state solution by setting $\partial \mathbf{s} / \partial t = 0$, allowing us to drop the time dependence from $\mathbf{s}$. The component parallel to the magnetic field decouples from the others, giving a pair of coupled equations for the in-plane spin density, written in matrix form as
\begin{align}
\frac{\text{d}^2}{\text{d}x^2}
\begin{bmatrix}
s_x \\
s_y
\end{bmatrix}
=
\begin{bmatrix}
1/\lambda^2 & -\omega/D \\
\omega/D & 1/\lambda^2
\end{bmatrix}
\begin{bmatrix}
s_x \\
s_y
\end{bmatrix}, \nonumber
\end{align}
where $\lambda = \sqrt{D\tau}$ is the spin diffusion length. Here we have assumed isotropic spin relaxation in the graphene plane, $\tau_\text{s}^x = \tau_\text{s}^y \equiv \tau$. The general solution to this system of equations is given by
\begin{align}
\begin{bmatrix}
s_x \\
s_y
\end{bmatrix}
=
\begin{bmatrix}
\text{i}\text{e}^{-x/\tilde{\lambda}} & \text{i}\text{e}^{x/\tilde{\lambda}} & -\text{i}\text{e}^{-x/\tilde{\lambda}^*} & -\text{i}\text{e}^{x/\tilde{\lambda}^*} \\
\text{e}^{-x/\tilde{\lambda}} & \text{e}^{x/\tilde{\lambda}} & \text{e}^{-x/\tilde{\lambda}^*} & \text{e}^{x/\tilde{\lambda}^*}
\end{bmatrix}
\begin{bmatrix}
Q \\ R \\ S \\ T
\end{bmatrix}, \nonumber
\end{align}
where $\tilde{\lambda} = \lambda / \sqrt{1+\text{i}\omega\tau}$ and $\{Q,R,S,T\}$ are coefficients determined by the boundary conditions of the particular system. These general expressions for $s_x$ and $s_y$ can now be applied to the case shown in Fig.\ \ref{fig_nonlocal}.

\subsection{Nonuniform Channel}

Now we consider the situation where part of the graphene channel is covered by a 2DFM. The graphene can be split into four regions, labeled 1-4 in Fig.\ \ref{fig_nonlocal}, and we assume that $D$, $\tau$, and $\omega$ (and thus $\lambda$) can be different in each region. The general expressions for $s_x$ and $s_y$ in each region then become
\begin{align}
\begin{bmatrix}
s_x^m \\
s_y^m
\end{bmatrix}
=
\begin{bmatrix}
\text{i}\text{e}^{-x/\tilde{\lambda}_m} & \text{i}\text{e}^{x/\tilde{\lambda}_m} & -\text{i}\text{e}^{-x/\tilde{\lambda}_m^*} & -\text{i}\text{e}^{x/\tilde{\lambda}_m^*} \\
\text{e}^{-x/\tilde{\lambda}_m} & \text{e}^{x/\tilde{\lambda}_m} & \text{e}^{-x/\tilde{\lambda}_m^*} & \text{e}^{x/\tilde{\lambda}_m^*}
\end{bmatrix}
\begin{bmatrix}
Q_m \\ R_m \\ S_m \\ T_m
\end{bmatrix}, \nonumber
\end{align}
where $m = 1,2,3,4$.

Next we define $x_m$ as the position of the interface between regions $m$ and $m+1$. The spin density and its derivative are continuous across each interface, except at the injector contact where we are injecting a spin current $I_\text{s}$ that is polarized along the $y$-axis,
\begin{align}
s_{x,y}^m(x_m) &= s_{x,y}^{m+1}(x_m) \nonumber \\
\left. \frac{\text{d} s_x^m}{\text{d} x} \right|_{x=x_m} &= \left. \frac{\text{d} s_x^{m+1}}{\text{d} x} \right|_{x=x_m} \nonumber \\
\left. \frac{\text{d} s_y^m}{\text{d} x} \right|_{x=x_m} &= \left. \frac{\text{d} s_y^{m+1}}{\text{d} x} \right|_{x=x_m}
+
\begin{Bmatrix}
\frac{I_\text{s}}{D}, m=1 \\ 0, m \neq 1
\end{Bmatrix}
. \nonumber
\end{align}

With these conditions, we arrive at the following set of equations at each interface,
\onecolumngrid
\begin{align}
\begin{bmatrix}
\text{e}^{-\frac{x_m}{\tilde{\lambda}_m}} & \text{e}^{\frac{x_m}{\tilde{\lambda}_m}} & -\text{e}^{-\frac{x_m}{\tilde{\lambda}_m^*}} & -\text{e}^{\frac{x_m}{\tilde{\lambda}_m^*}} & -\text{e}^{-\frac{x_m}{\tilde{\lambda}_{m+1}}} & -\text{e}^{\frac{x_m}{\tilde{\lambda}_{m+1}}} & \text{e}^{-\frac{x_m}{\tilde{\lambda}_{m+1}^*}} & \text{e}^{\frac{x_m}{\tilde{\lambda}_{m+1}^*}}
\\
\text{e}^{-\frac{x_m}{\tilde{\lambda}_m}} & \text{e}^{\frac{x_m}{\tilde{\lambda}_m}} & \text{e}^{-\frac{x_m}{\tilde{\lambda}_m^*}} & \text{e}^{\frac{x_m}{\tilde{\lambda}_m^*}} & -\text{e}^{-\frac{x_m}{\tilde{\lambda}_{m+1}}} & -\text{e}^{\frac{x_m}{\tilde{\lambda}_{m+1}}} & -\text{e}^{-\frac{x_m}{\tilde{\lambda}_{m+1}^*}} & -\text{e}^{\frac{x_m}{\tilde{\lambda}_{m+1}^*}}
\\
\frac{\text{e}^{-\frac{x_m}{\tilde{\lambda}_m}}}{\tilde{\lambda}_m} & -\frac{\text{e}^{\frac{x_m}{\tilde{\lambda}_m}}}{\tilde{\lambda}_m} & -\frac{\text{e}^{-\frac{x_m}{\tilde{\lambda}_m^*}}}{\tilde{\lambda}_m^*} & \frac{\text{e}^{\frac{x_m}{\tilde{\lambda}_m^*}}}{\tilde{\lambda}_m^*} & -\frac{\text{e}^{-\frac{x_m}{\tilde{\lambda}_{m+1}}}}{\tilde{\lambda}_{m+1}} & \frac{\text{e}^{\frac{x_m}{\tilde{\lambda}_{m+1}}}}{\tilde{\lambda}_{m+1}} & \frac{\text{e}^{-\frac{x_m}{\tilde{\lambda}_{m+1}^*}}}{\tilde{\lambda}_{m+1}^*} & -\frac{\text{e}^{\frac{x_m}{\tilde{\lambda}_{m+1}^*}}}{\tilde{\lambda}_{m+1}^*}
\\
\frac{\text{e}^{-\frac{x_m}{\tilde{\lambda}_m}}}{\tilde{\lambda}_m} & -\frac{\text{e}^{\frac{x_m}{\tilde{\lambda}_m}}}{\tilde{\lambda}_m} & \frac{\text{e}^{-\frac{x_m}{\tilde{\lambda}_m^*}}}{\tilde{\lambda}_m^*} & -\frac{\text{e}^{\frac{x_m}{\tilde{\lambda}_m^*}}}{\tilde{\lambda}_m^*} & -\frac{\text{e}^{-\frac{x_m}{\tilde{\lambda}_{m+1}}}}{\tilde{\lambda}_{m+1}} & \frac{\text{e}^{\frac{x_m}{\tilde{\lambda}_{m+1}}}}{\tilde{\lambda}_{m+1}} & -\frac{\text{e}^{-\frac{x_m}{\tilde{\lambda}_{m+1}^*}}}{\tilde{\lambda}_{m+1}^*} & \frac{\text{e}^{\frac{x_m}{\tilde{\lambda}_{m+1}^*}}}{\tilde{\lambda}_{m+1}^*}
\end{bmatrix}
\begin{bmatrix}
Q_m \\ R_m \\ S_m \\ T_m \\ Q_{m+1} \\ R_{m+1} \\ S_{m+1} \\ T_{m+1}
\end{bmatrix}
=
\begin{bmatrix}
0 \\ 0 \\ 0 \\
\begin{Bmatrix}
\frac{I_\text{s}}{D}, m=1 \\ 0, m \neq 1
\end{Bmatrix}
\end{bmatrix}. \nonumber
\end{align}
\twocolumngrid

Applying this set of equations at each interface ($m = 1, 2, 3$) gives us 12 equations for 16 unknowns. Next we assume the spin density goes to 0 as $x \rightarrow \pm \infty$; this is equivalent to saying that any reference contacts or edges of the graphene flake are far enough away from the injector/detector that they play no role in the measured spin signal. This assumption allows us to set $Q_1 = S_1 = R_4 = T_4 = 0$, leaving us with 12 equations for 12 unknowns. After some algebra, we arrive at the following expression for the nonlocal voltage measured at the detector contact,
\begin{align} \label{hanle_nonuniform}
V_\text{NL} &\propto s_y(x = L_\text{ch}) \nonumber \\
&= \frac{I_\text{s}}{D} \cdot 2\operatorname{Re}\left\{ \frac{\tilde{\lambda} \text{e}^{-(L_\text{ch}-L_\text{H})/\tilde{\lambda}}}{ \frac{(\tilde{\lambda}_\text{H}+\tilde{\lambda})^2}{\tilde{\lambda}_\text{H} \tilde{\lambda}} \text{e}^{L_\text{H}/\tilde{\lambda}_\text{H}} - \frac{(\tilde{\lambda}_\text{H}-\tilde{\lambda})^2}{\tilde{\lambda}_\text{H} \tilde{\lambda}} \text{e}^{-L_\text{H}/\tilde{\lambda}_\text{H}}} \right\},
\end{align}
where $\tilde{\lambda}_\text{H} = \lambda_\text{H} / \sqrt{1+\text{i}\omega_\text{H}\tau_\text{H}}$, $\lambda_\text{H} = \sqrt{D_\text{H}\tau_\text{H}}$, and $\omega_\text{H} = \gamma(B+B_\text{H})$.

Equation (\ref{hanle_nonuniform}) is the main result of this work, and can be used to fit measurements of Hanle spin precession in the setup shown in Fig.\ \ref{fig_nonlocal}. Before exploring the features of the Hanle precession in the next section, we quickly point out that by setting $\tilde{\lambda}_\text{H} = \tilde{\lambda}$ we recover the usual expression for the uniform channel,
\begin{align} \label{hanle_uniform}
s_y(x = L_\text{ch}) = \frac{I_\text{s}}{2D} \operatorname{Re}\left\{ \tilde{\lambda} \text{e}^{-L_\text{ch}/\tilde{\lambda}} \right\}.
\end{align}
Comparison to Eq.\ (\ref{hanle_uniform}) reveals that the numerator of Eq.\ (\ref{hanle_nonuniform}) describes the uncovered portion of the graphene, renormalized by a denominator describing the graphene/2DFM region.

\section{Features of the Model} \label{sec_model_features}

\begin{figure}[t]
\includegraphics[width=\columnwidth]{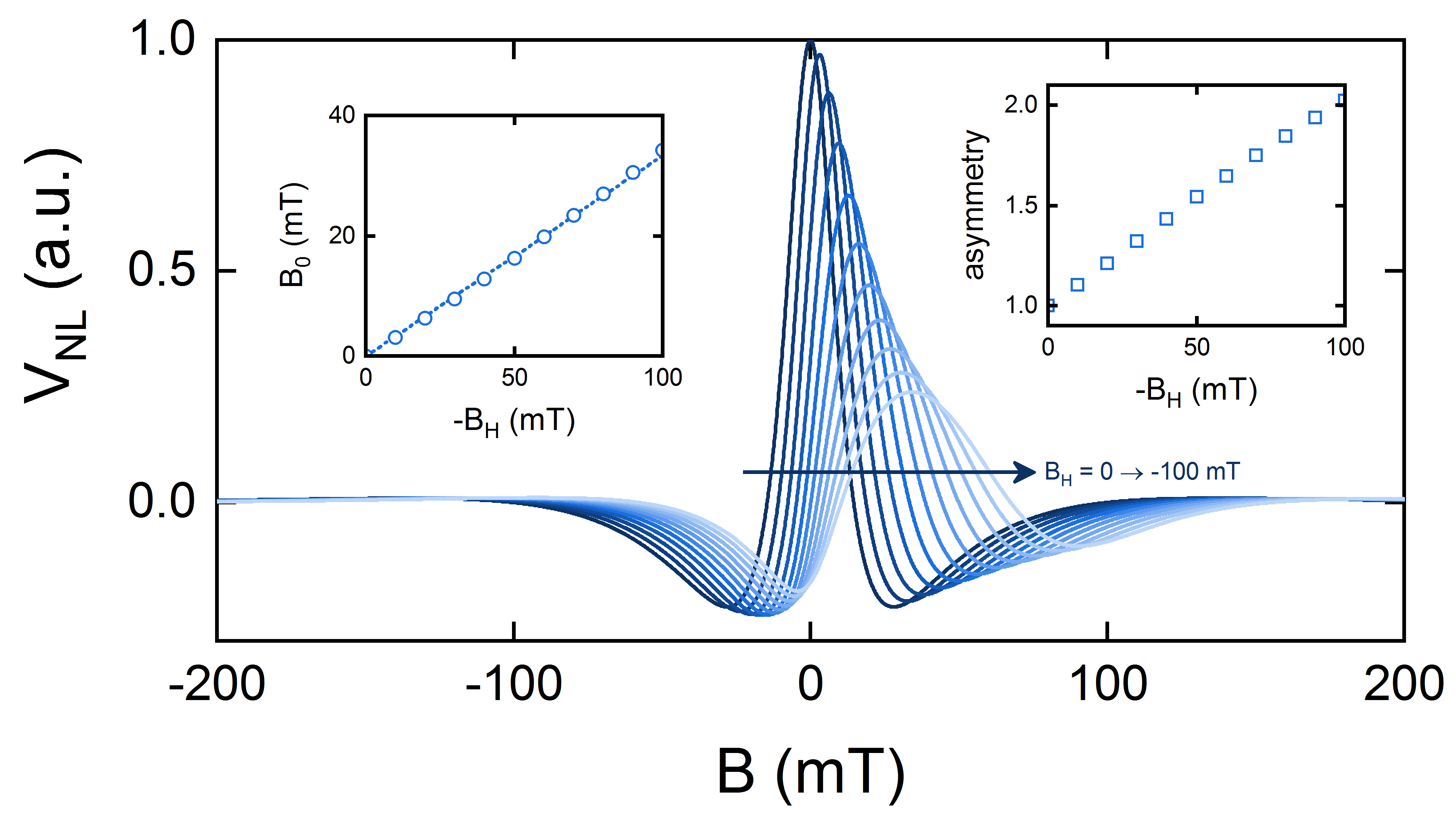}
\caption{Impact of proximity-induced exchange field. Simulation parameters are $L_\text{ch} = 10$ $\upmu$m, $L_\text{H} = 5$ $\upmu$m, $\tau_\text{H} = \tau = 500$ ps, $D_\text{H} = D = 0.05$ m$^2$/s, and $B_\text{H}$ is varied from $0 \rightarrow -100$ mT in steps of $-10$ mT. The left inset shows the shift of the Hanle peak with increasing $B_\text{H}$, where the symbols are extracted from the Hanle curves in the main panel and the dashed line is Eq.\ (\ref{eq_beff}). The right inset shows the ratio of the left and right minima.}
\label{fig_hanle_Bh}
\end{figure}

To investigate the qualitative features of this model, we start from a typical set of experimental parameters for graphene nonlocal spin valves: $L_\text{ch} = 10$ $\upmu$m, $\tau = 500$ ps, and $D = 0.05$ m$^2$/s. We then vary $L_\text{H}$ and $\tau_\text{H}$, and we assume $D_\text{H} = D$. To keep things simple and demonstrate the main features, we assume that $B_\text{H}$ is constant, while in an experimental situation it may vary with applied magnetic field $B$. In the first case we let the 2DFM cover half of the channel, $L_\text{H} = 5$ $\upmu$m, and we assume uniform spin transport, $\tau_\text{H} = \tau$. Figure \ref{fig_hanle_Bh} shows the sequence of Hanle curves, calculated with Eq.\ (\ref{hanle_nonuniform}), as the proximity-induced exchange field $B_\text{H} = 0 \rightarrow -100$ mT in steps of $-10$ mT. The magnitudes of the curves are normalized to the peak magnitude at $B_\text{H} = 0$.

In Fig.\ \ref{fig_hanle_Bh} we can see three main features. First, the peak of the Hanle curve shifts to the right with increasing $|B_\text{H}|$. This shift is the strength of the external magnetic field $B = B_0$ needed to cancel out $B_\text{H}$, resulting in zero net precession of spins as they traverse the channel. The symbols in the left inset show a linear relationship between $B_0$ and $B_\text{H}$, extracted from the Hanle curves in the main panel. This relationship is well-captured by the expression
\begin{align} \label{eq_beff}
B_0 \approx B_\text{H} \cdot \frac{L_\text{H}}{L_\text{ch} - (1 - \frac{\lambda}{\lambda_\text{H}})L_\text{H} + \lambda},
\end{align}
as shown by the dashed line in the inset. Equation (\ref{eq_beff}) accounts for the partial coverage of the graphene by the 2DFM and describes the total effective magnetic field felt by electrons as they traverse the channel. The second feature is a decrease of the magnitude of the central Hanle peak with increasing $|B_\text{H}|$. As the magnetic field needed to cancel $B_\text{H}$ increases, the dephasing induced by spin precession in the uncovered region also increases, leading to a reduced spin signal. This also explains the third main feature, which is an asymmetry in the minima of the curve on each side of the central peak. These minima correspond to a net 180$^\text{o}$ rotation of the spins as they traverse the graphene channel. On the positive side of the central peak, the magnitude of $B$ needed to induce a net 180$^\text{o}$ rotation is larger than that needed on the negative side because $B_\text{H} < 0$. This leads to stronger dephasing and a smaller minimum on the positive side compared to the negative side. The ratio of these two minima is shown in the right inset of Fig.\ \ref{fig_hanle_Bh}, demonstrating increasing asymmetry with increasing $|B_\text{H}|$.

\begin{figure}[t]
\includegraphics[width=\columnwidth]{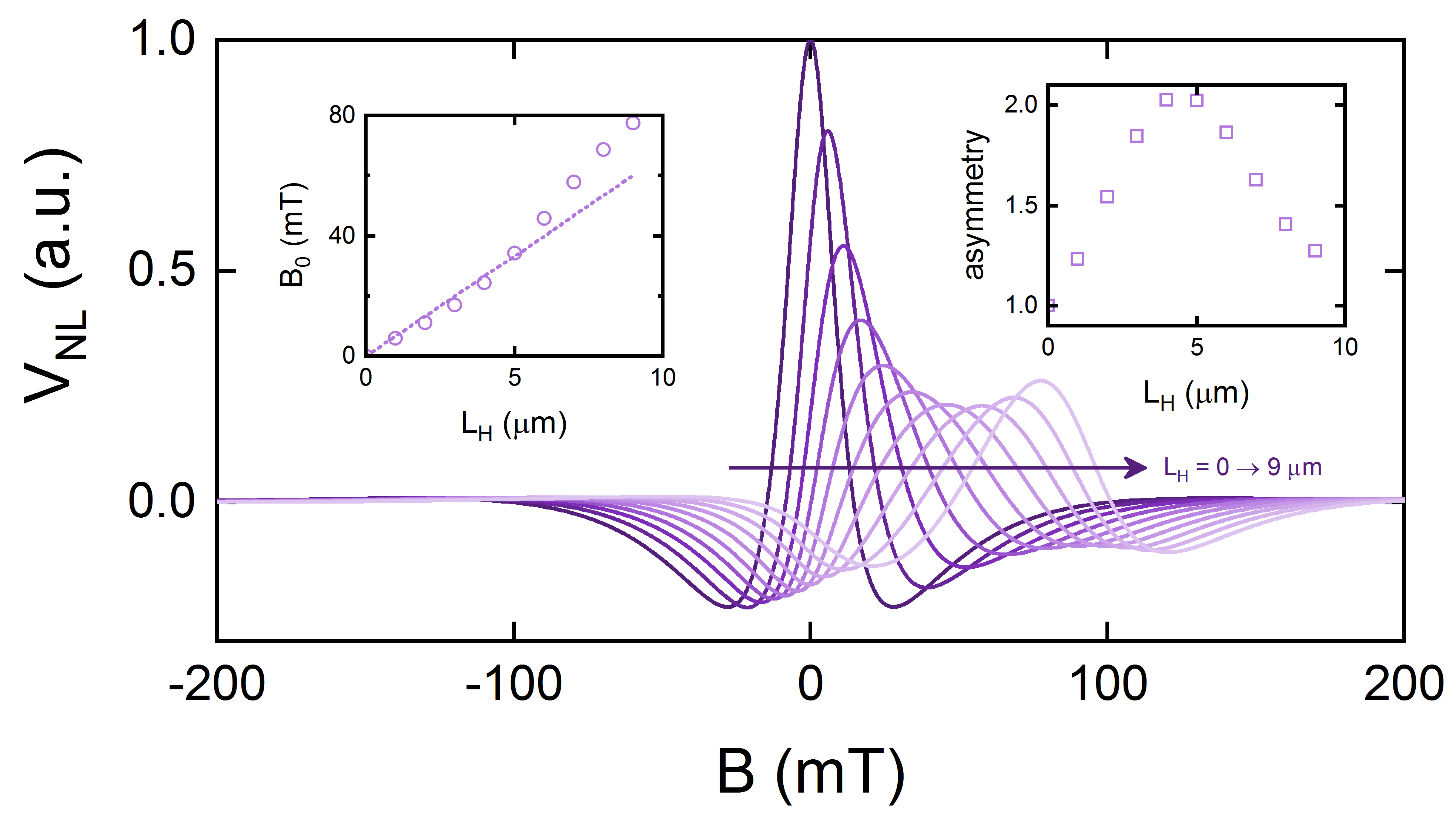}
\caption{Impact of graphene/2DFM overlap. Simulation parameters are $L_\text{ch} = 10$ $\upmu$m, $B_\text{H} = -100$ mT, $\tau_\text{H} = \tau = 500$ ps, $D_\text{H} = D = 0.05$ m$^2$/s, and $L_\text{H}$ is varied from $0 \rightarrow 9$ $\upmu$m in steps of $1$ $\upmu$m. The left inset shows the shift of the Hanle peak with increasing $L_\text{H}$, where the symbols are extracted from the Hanle curves in the main panel and the dashed line is Eq.\ (\ref{eq_beff}). The right inset shows the ratio of the left and right minima.}
\label{fig_hanle_Lh}
\end{figure}

Next we fix $B_\text{H} = -100$ mT and we vary the coverage of the 2DFM region, $L_\text{H} = 0 \rightarrow 9$ $\upmu$m in steps of 1 $\upmu$m. This is shown in Fig.\ \ref{fig_hanle_Lh}. Similar to before, the shift of the Hanle peak increases with increasing graphene/2DFM coverage, and is also well-described by Eq.\ (\ref{eq_beff}). For small values of $L_\text{H}$, the magnitude of the central Hanle peak decreases as 2DFM coverage increases, but as $L_\text{H} \rightarrow L_\text{ch}$ this decrease slows down and reverses. At large coverage, the effective magnetic field at the Hanle peak, $B_\text{H} + B_0$, is almost completely uniform and close to 0, with only a small uncovered portion that sees a large external $B$. This corresponds to reduced spin dephasing and a larger Hanle signal when approaching full coverage. Similar behavior is seen in the asymmetry of the minima in the Hanle curve; the asymmetry initially grows, then slows down and reverses as the graphene channel becomes more uniformly covered by the 2DFM.

\begin{figure}[t]
\includegraphics[width=\columnwidth]{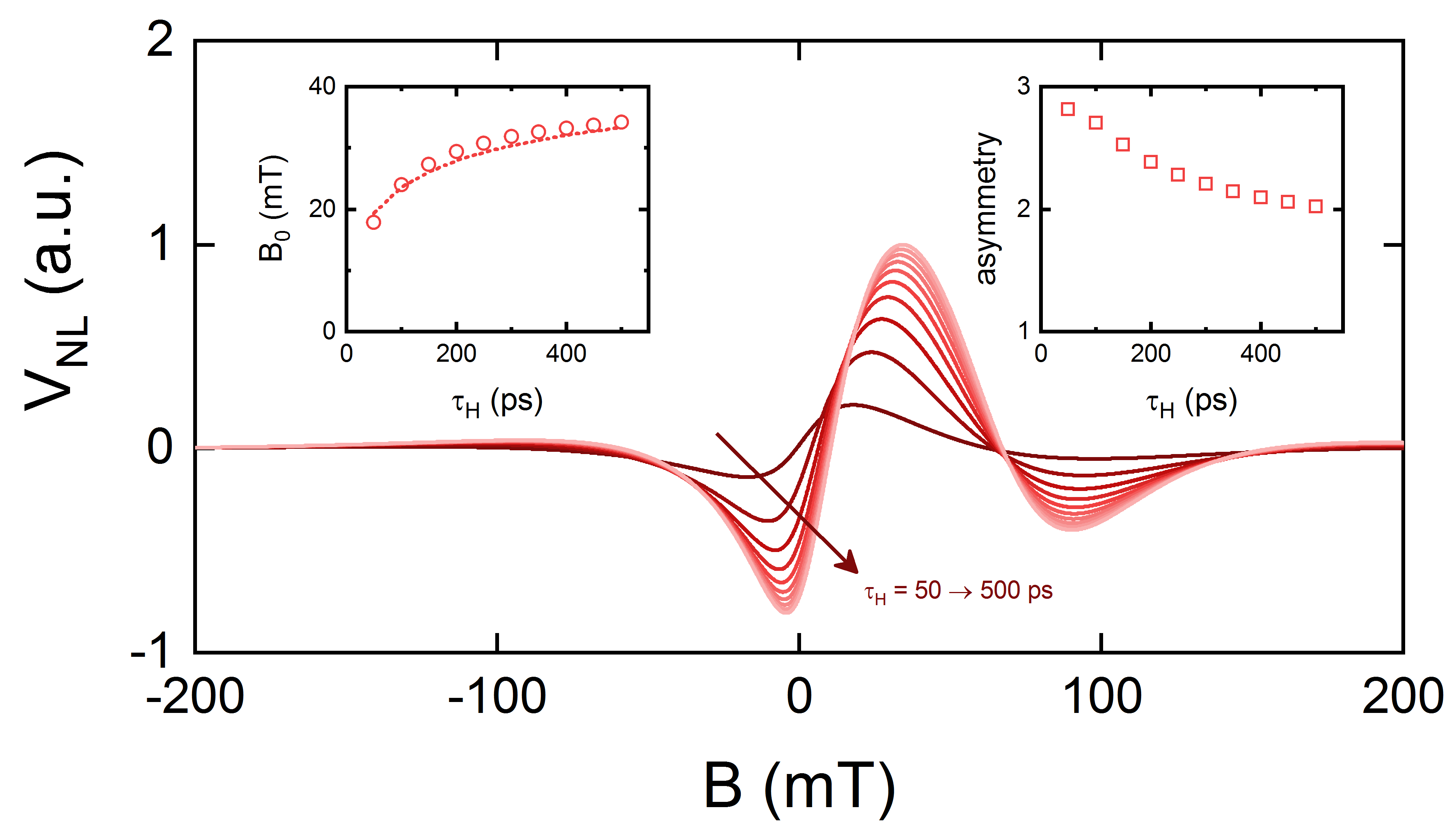}
\caption{Impact of graphene/2DFM spin lifetime. Simulation parameters are $L_\text{ch} = 10$ $\upmu$m, $L_\text{H} = 5$ $\upmu$m, $B_\text{H} = -100$ mT, $\tau = 500$ ps, $D_\text{H} = D = 0.05$ m$^2$/s, and $\tau_\text{H}$ is varied from $50 \rightarrow 500$ ps in steps of $50$ ps. The left inset shows the shift of the Hanle peak with increasing $\tau_\text{H}$, where the symbols are extracted from the Hanle curves in the main panel and the dashed line is Eq.\ (\ref{eq_beff}). The right inset shows the ratio of the left and right minima.}
\label{fig_hanle_th}
\end{figure}

Finally, in Fig.\ \ref{fig_hanle_th} we fix $B_\text{H} = -100$ mT and $L_\text{H} = 5$ $\upmu$m, and we vary the spin relaxation time in the graphene/2DFM region, $\tau_\text{H} = 50 \rightarrow 500$ ps in steps of $50$ ps. Here, the magnitude of the Hanle signal decreases with decreasing $\tau_\text{H}$, a result of overall faster spin relaxation. The shift of the Hanle peak also decreases with decreasing $\tau_\text{H}$, as the spins in the graphene/2DFM region relax before they can be rotated by the proximity-induced exchange field $B_\text{H}$. This shift is also well-captured by Eq.\ (\ref{eq_beff}), as shown in the left inset.

\section{Spurious Effects} \label{sec_spurious}

In the previous section we saw that two main features arise in the Hanle spin precession of graphene in proximity to a 2DFM. First, there is a shift of the Hanle peak due to the proximity-induced exchange field $B_\text{H}$. Second, there is an asymmetry in the Hanle curve that arises from the finite coverage of the graphene channel by the 2DFM. In this section we show that these features are not unique, as they can also arise from misaligned injector/detector contacts.

Consider the situation shown in Fig.\ \ref{fig_nonlocal}, where the detector contact is rotated by an angle $\alpha$ with respect to the injector contact. In this scenario, the measured nonlocal voltage $V_\text{NL}$ will consist of both the $x$ and $y$ components of the spin density at the detector,
\begin{align} \label{}
V_\text{NL} &\propto \cos(\alpha) s_y(x = L_\text{ch}) + \sin(\alpha) s_x(x = L_\text{ch}) \nonumber \\
&= \frac{I_\text{s}}{D} \cdot 2 \left[\cos(\alpha)\operatorname{Re}\{ ... \} - \sin(\alpha)\operatorname{Im}\{ ... \} \right], \nonumber
\end{align}
where $\{ ... \}$ refers to the expression in the brackets in Eq.\ (\ref{hanle_nonuniform}). The $\operatorname{Re}\{ ... \}$ and $\operatorname{Im}\{ ... \}$ terms respectively give rise to a cosine-like and sine-like signal as a function of $B$, and their sum can yield an apparent shift of the Hanle peak, as well as asymmetry in the Hanle minima \cite{Ringer2018}.

\begin{figure}[H]
\includegraphics[width=\columnwidth]{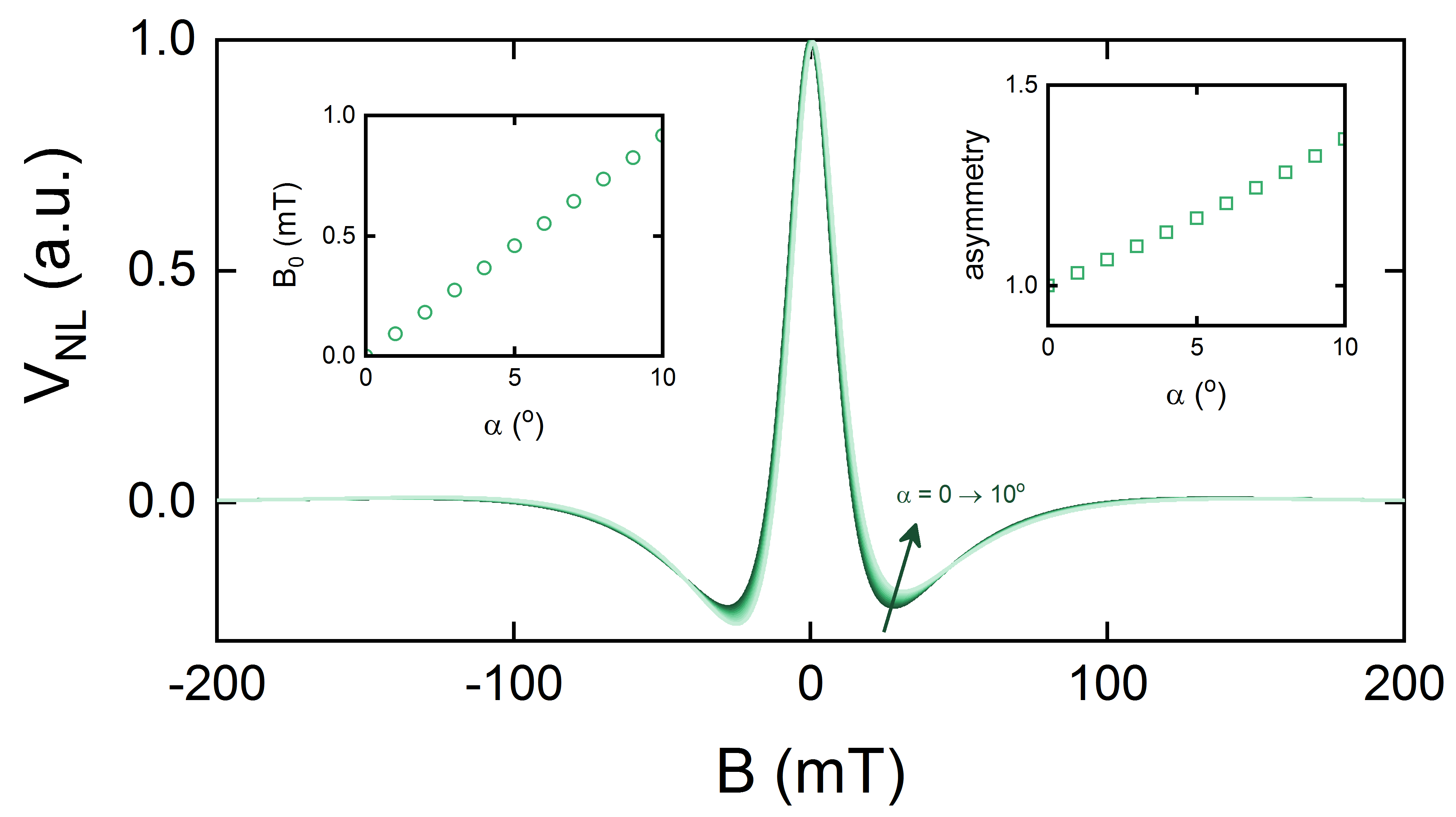}
\caption{Impact of detector contact rotation. Simulation parameters are $L_\text{ch} = 10$ $\upmu$m, $L_\text{H} = 5$ $\upmu$m, $B_\text{H} = 0$, $\tau_\text{H} = \tau = 500$ ps, $D_\text{H} = D = 0.05$ m$^2$/s, and we vary the detector contact rotation $\alpha = 0 \rightarrow 10^\text{o}$ in steps of $1^\text{o}$. The left inset shows the shift of the Hanle peak with increasing $\alpha$, and the right inset shows the ratio of the left and right minima.}
\label{fig_hanle_alpha}
\end{figure}

In Fig.\ \ref{fig_hanle_alpha} we show how rotation of the detector contact impacts the Hanle signal. We take the same set of parameters as above, $L_\text{ch} = 10$ $\upmu$m, $L_\text{H} = 5$ $\upmu$m, $B_\text{H} = 0$, $\tau_\text{H} = \tau = 500$ ps, $D_\text{H} = D = 0.05$ m$^2$/s, and we vary $\alpha = 0 \rightarrow 10^\text{o}$ in steps of $1^\text{o}$. Although the effect is smaller in this case, we can see that increasing detector misalignment leads to an apparent shift of the Hanle peak (shown in the left inset), as well as increasing asymmetry of the Hanle minima (right inset).

\section{Summary and Discussion} \label{sec_experiment}

In this paper we have derived an expression, Eq.\ (\ref{hanle_nonuniform}), that describes Hanle spin precession in graphene interfaced with a 2DFM that exhibits perpendicular magnetic anisotropy. This expression reveals two signatures when the 2DFM is ferromagnetic: a shift of the Hanle peak, and asymmetry in the Hanle curve. For the situations studied above, the shift of the Hanle peak $B_0$ is approximately related to the device parameters according to Eq.\ (\ref{eq_beff}). This relation accounts for the partial coverage of the channel by the 2DFM, as well as different spin lifetime under the 2DFM. Meanwhile, the asymmetry of the Hanle curve is a consequence of the finite extent of the graphene/2DFM interface region, with maximal asymmetry when the 2DFM covers $\sim$50\% of the channel. These two features have been seen in recent measurements of graphene/CGT heterostructures \cite{Karpiak2019}, and should be general to graphene interfaced with any material exhibiting out-of-plane ferromagnetism. We therefore hope that this analysis and Eq.\ (\ref{hanle_nonuniform}) will be useful for future studies of these types of heterostructures.

We would like to point out that when using this expression to analyze experimental results, a couple considerations should be made. First, Eq.\ (\ref{hanle_nonuniform}) contains a large number of parameters, and it may be difficult to extract a unique fit to an experimental Hanle curve. For this reason a control device consisting of uncovered graphene can be used to first extract values for $\tau$ and $D$, leaving $\tau_\text{H}$, $D_\text{H}$, and $B_\text{H}$ as the only fitting parameters. Second, as shown in Sec.\ \ref{sec_spurious}, contact misalignment can result in a signal that mimics the presence of a perpendicular exchange field. By employing temperature-dependent measurements of Hanle spin precession, one should be able to disentangle these two effects as long as the contacts and the 2DFM have different Curie temperatures.

Finally, this work has focused on the case of graphene interfaced with 2DFM insulators, but in principle it can be applied to a variety of other magnetic systems. For example, the recent discovery that twisted bilayer graphene can be superconducting \cite{Cao2018} has generated a significant amount of interest in twisted layered systems. One example is twisted double bilayer graphene (TDBG), which appears to exhibit electrically-tunable ferromagnetism \cite{Liu2019}. The strength of this effect could be studied with the setup of Fig.~\ref{fig_nonlocal} and with Eq.~(\ref{hanle_nonuniform}), where the graphene/2DFM region is replaced by a TDBG stack.

\begin{acknowledgments}
ICN2 is supported by the Severo Ochoa program from Spanish MINECO (grant no.\ SEV-2017-0706) and funded by the CERCA Programme / Generalitat de Catalunya. AWC acknowledges support from the European Union Horizon 2020 Programme under grant agreement no.\ 785219 (Graphene Flagship Core 2), as well as comments from S.\ Roche, J.H. Garcia, M.\ Vila, and the referees.
\end{acknowledgments}

\clearpage
\bibliography{Hanle_graphene_2dfm_bib}

\end{document}